\documentclass[
 	reprint,
	amsmath,
    amssymb,
	aps,
	longbibliography,
    floatfix,
]{revtex4-1}

\usepackage[english]{babel}
\usepackage[utf8x]{inputenc}
\usepackage[T1]{fontenc}
\usepackage{braket}

\usepackage{amsmath}
\usepackage{graphicx}
\usepackage{dcolumn}
\usepackage{bm}
\usepackage{array}
\usepackage[shortlabels]{enumitem}
\newcolumntype{?}{!{\vrule width 1pt}}
\usepackage[colorinlistoftodos]{todonotes}
\usepackage[colorlinks=true, allcolors=blue]{hyperref}
\newcommand{\angstrom}{\textup{\AA}}

\makeatletter
\let\cat@comma@active\@empty
\makeatother

\begin{document} 
\title{Understanding disorder in 2D materials: the case of carbon doping of Silicene} 
\author{Ricardo Pablo-Pedro$^{1,2}$, Miguel Angel Maga\~na-Fuentes$^{3}$, Marcelo Videa$^{4}$, Jing Kong$^{5}$, Mingda Li$^{2}$, Jose L. Mendoza-Cortes$^{3,6,7,*}$, and Troy Van Voorhis$^{1,*}$\\ \hspace{0.1cm}} 
\affiliation{\\
$^{1}$Department of Chemistry, MIT, 77 Massachusetts Avenue, Cambridge, MA 02139, USA\\
$^{2}$Department of Nuclear Science and Engineering, MIT, Cambridge, Massachusetts 02139, USA\\
$^{3}$Department of Chemical \& Biomedical Engineering, FAMU-FSU joint College of Engineering, 2525 Pottsdamer St, Tallahassee, FL 32310, USA\\
$^{4}$School of Engineering and Sciences, Tecnologico de Monterrey, Campus Monterrey, Av. Eugenio Garza Sada 2501 Sur. Monterrey N. L., Monterrey 64849, Mexico\\
$^{5}$Department of Electrical Engineering and Computer Science, MIT, Cambridge, MA 02139, USA\\
$^{6}$ Department of Physics, Scientific Computing Department, Materials Science and Engineering, High Performance Material Institute, Condensed Matter Theory - National High Magnetic Field Laboratory, Florida State University, 77 Chieftan Way, Tallahassee, FL 32306, USA\\
$^{7}$ \textit{Current Address:} Department of Chemical Engineering \& Materials Science, Michigan State University, East Lansing, Michigan 48824, United States.\\
\newline 
$^{*}$Corresponding authors: jmendoza@msu.edu and tvan@mit.edu. \\
}


\begin{abstract}
We investigate the effect of lattice disorder and local correlation effects in finite and periodic silicene structures caused by carbon doping using first-principles calculations. For both finite and periodic silicene structures, the electronic properties carbon-doped monolayers are dramatically changed by controlling the doping sites in the structures, which is related to the amount of disorder introduced in the lattice and electron-electron correlation effects. By changing the position of the carbon dopants, we found that a Mott-Anderson transition is achieved. Moreover, the band gap is determined by the level of lattice disorder and electronic correlation effects. Finally, these structures are ferromagnetic even under disorder which has potential applications in Si-based nanoelectronics, such as field-effect transistors (FETs).      
\end{abstract}

\maketitle 

From the large variety of two-dimensional (2D) materials that exist today, silicene, the silicon counterpart of graphene has steadily increased its sphere of influence due to its malleable electronic properties and its compatibility with the current silicon-based technology.\cite{ZHAO201624} For instance, the buckled layer geometry of silicene facilitates a band gap engineering and in the presence of an electric field opens a gap transport that makes possible the realization of a field-effect transistor (FET) at room temperature.\cite{doi:10.1021/nl203065e} Synthesis of silicene is achieved nowadays by surface-assisted epitaxial growth on different substrates;\cite{Lalmi_2010, MengIr_2013, Fleurence_2012} however, during this process, the formation of imperfections on the layer is practically unavoidable, which strongly influences the magnetic and electronic properties of the material. \cite{guzman2007electronic,cahangirov2009two,kara2012review,lin2016defect,Li_def_2015} In the context of imperfections, these modifications of the electronic, magnetic, and structural properties imply a fundamental form of disorder.\cite{rhodes2019disorder}

\begin{figure*}[ht!]
  \includegraphics[width=17cm]{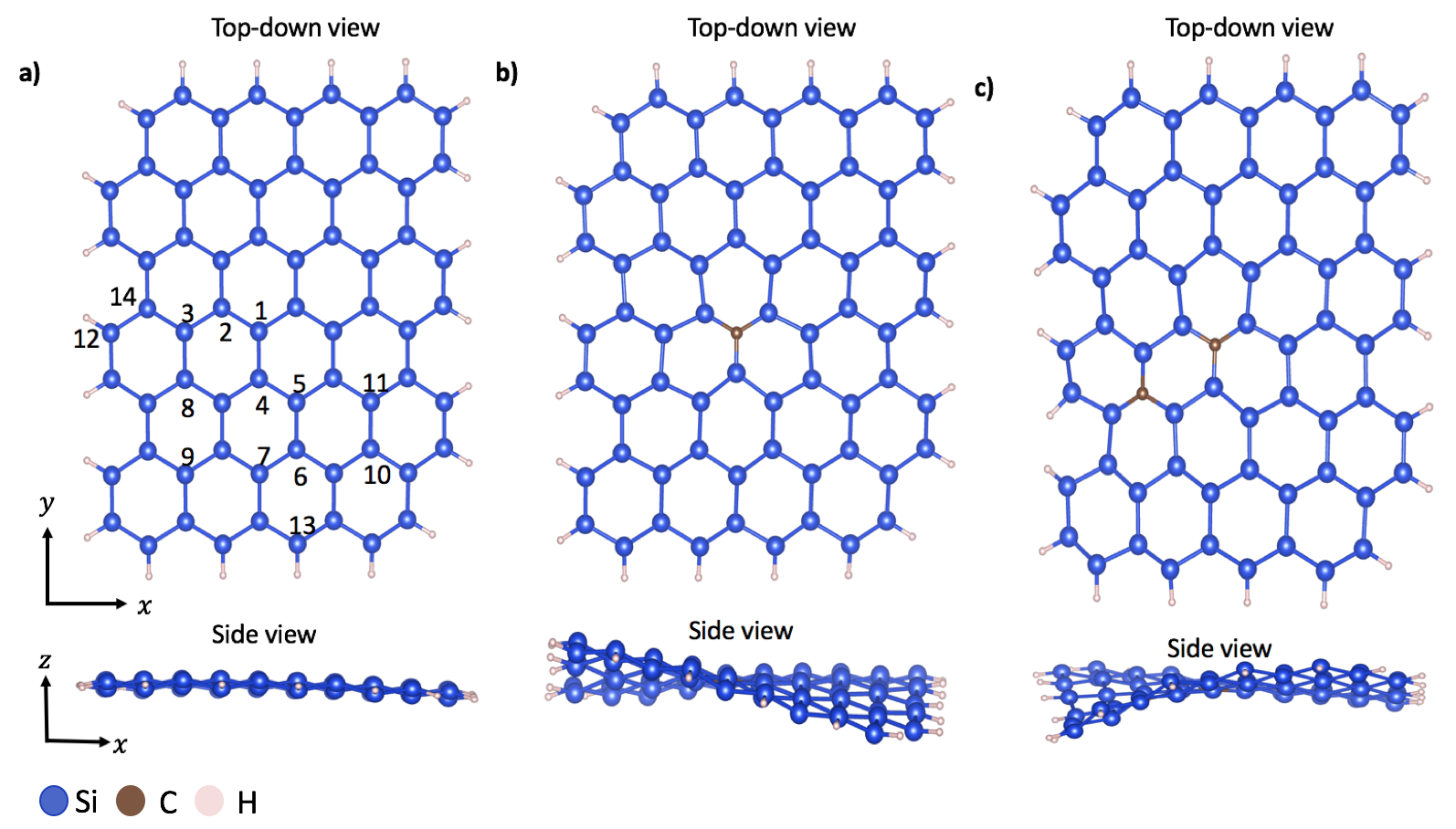}
  \caption{{\bfseries Equilibrium structures of pure, single- and double-doped silicene nanocluster.} {\bfseries a)} Pure silicene nanocluster, which has fourteen nonequivalent possible doping positions. {\bfseries b)} Silicene single-doped with carbon generates out-of-plane distortion. {\bfseries c)} Double-doped carbon substitution generates both out-of-plane and bonds distortions.}
  \label{fig:silicenedopanscheme}
\end{figure*}

To illustrate the origin and impact of disorder in silicene, we use chemical doping substitution since it is the simplest way of introducing disorder and of modifying 2D materials' electronic and structural properties beyond  electrically-controlled means.\cite{chen2008charged,sachs2013doping}
Unlike Graphene, Silicene's monolayers are also particularly sensitive to alterations in the geometry of the lattice due to their inherent buckling structure; moreover, Silicene's monolayers present two parallel sublattices which induces symmetry breaking between them when defects are introduced. This symmetry between the sublattices produces a position-dependence when more than one defect is introduced. Different arrangements have distinct electronic properties. These geometry and symmetry considerations make Silicene prone to a diverse range of phase transitions. Depending on the concentration, distribution, and nature of the doping atoms, the monolayers undergo semimetal to semiconductor or metal transitions.

Consider, for instance, the case of phosphorus-doped silicon (3D structure), a particularly well-studied example of a system showing a metal-insulator transition (MIT).\cite{PhysRevB.27.7509} According to Mott's argument, a transition from an insulator to metal occurs at a critical concentration, $n_{c}$.\cite{RevModPhys.40.677} Although a temperature, $T=0$, is necessary for the MIT, several physical systems have shown an MIT at finite temperatures. \cite{anissimova2007flow,lilly2003resistivity,limelette2003mott} Hence the importance of studying this transition. A MIT  which is driven mostly by the electronic correlations is known  Mott-Hubbard transition;\cite{zwerger2003mott} while an Anderson transition is driven mostly by lattice disorder.\cite{RevModPhys.80.1355}

The semimetallic character of silicene limits its potential as a suitable material for distinct applications; however, this limitation could be overcome by the induction of Mott-Hubbard or Anderson transitions. Various authors have studied in detail silicene-graphene hybrid layers and their properties;\cite{Shi_SiC_2015, Vahabzadeh_2019} however, to the best of our knowledge, there is no systematic study on the behavior of carbon-doped silicene in simulations or experiments. In this work, we study the doping behavior of both finite and periodic structures of silicene with carbon atoms. We show that the electronic properties of carbon-doped silicene structures depend on disorder effects and electron-electron correlations. Also, we found that c doped periodic structures are more stable than monovacancies. Furthermore, we compare the electronic and structural properties of periodic and finite-size structures to have a better understanding of the influence of dopant atoms on the nanoclusters.

{\bf AB Initio Characterization}. For finite-size structures, we choose the pairing Beck 3-Parameter (exchange), Lee, Yang and Parr (B3LYP) functional with the correlation consistent polarized valence double-$\zeta$ (cc-pVDZ) base along with D3 dispersion correction,~\cite{Grimme2010} as implemented in Q-Chem 5.0.~\cite{shao2015advances} For the periodic structures, the numerical simulations were carried out using plane-wave basis as  implemented in the Vienna \textit{ab initio} simulation package (VASP).~\cite{PhysRevB.47.558,PhysRevB.49.14251,KRESSE199615,PhysRevB.54.11169,monkhorst1976special} Kinetic energy cutoff for plane-waves is set to 500 eV after convergence tests. B3LYP and Perdew-Burke-Ernzerhof (PBE) functionals were used for the periodic structure calculations to compare our results. For the PBE functional, the generalized gradient approximation (GGA) is carried out. 
 
\textbf{Results and discussions.} First, we study the geometrical and electronic properties of rectangular silicene nanocluster with mono-hydrogen termination at its edges (H-SiNC), see Fig.~\ref{fig:silicenedopanscheme}a). The selection of these specific structure was based on ref.~\cite{Pablo-Pedro2017}, which states that finite rectangular silicene nanoclusters with armchain edge lenght ($L_a$) larger than zigzag edge lenght ($L_z$) could only present ferromagnetic states that combined with silicon-based semiconductor technology could be valuable for spintronic applications. Different shapes could form different electronic states, see the supplementary material (SM) for more details. Here, we selected fourteen inequivalent doping positions $S_1$-$S_{14}$ for a single C atom; then, to study the stability of the doped structures, the defect formation energy, $E_f$, per dopant is calculated using the following equation   

\begin{equation}
\label{Eq.formationenery}
E_f=E_d-E_{ud}+nE_{Si}-nE_C
\end{equation}

\begin{table}\small
\caption{Formation energies (in eV per impurity atom) of a single  C substitution at thirteen different sites of H-SiNC for  singlet (S) and triplet (T) states at the B3LYP level of calculation. All formation energies were calculated with respect to the singlet undoped ground state of H-SiNC(4,6).}
\begin{tabular}{ccc}\\
    \toprule
Site & 1-Carbon (S)  & 1-Carbon(T)  \\
\hline
 $1$ & $-2.266$ & $-2.421$  \\ 
 $2$ & $-2.331$ & $-2.468$ \\
 $3$ & $-2.456$ & $-2.585$  \\
 $4$ & $-2.271$ & $-2.422$  \\
 $5$ & $-2.294$ & $-2.431$  \\
 $6$ & $-2.327$ & $-2.498$  \\
 $7$ & $-2.526$ & $-2.660$ \\
 $8$ & $-2.468$ & $-2.607$ \\
 $9$ & $-2.600$ & $-2.748$ \\
 $10$& $-2.532$ & $-2.681$  \\
 $11$& $-2.474$ & $-2.614$  \\
 $12$& $-3.372$ & $-3.495$ \\
 $13$& $-3.233$ & $-3.353$ \\
 $14$& $-2.849$ & $-2.997$ \\
   \hline
    \end{tabular}
 \label{table1}
\end{table}

\noindent where $E_d$ is the total energy of a doped structure, $E_{ud}$ is  the energy of a pristine structure, $n$ is the number of dopants, and the $E_{Si}$ and $E_C$ represent the energy of the free $Si$ and $C$ atoms.
The formation energies for single carbon substitutions are listed in Table \ref{table1}. Here, the singlet (antiferromagnetic) and triplet (ferromagnetic) states are reported since C-doped structures can be present in either state. The negative formation energies show that C doped systems are thermodynamically stable. For a single C substitution, the formation energy increases for sites $S_1$, $S_4$, $S_5$, $S_6$, $S_2$, $S_3$, $S_8$, $S_{11}$, $S_7$, $S_{10}$, $S_{14}$, $S_{13}$, and $S_{12}$, respectively. This shows that Si atoms closer to the edge are more likely to be substituted than those in the interior sites around the zigzag or armchair directions. Similar conclusions were obtained for  N and B doped silicene nanoribbons.~\cite{MA201366} Figures \ref{fig:silicenedopanscheme}b and \ref{fig:silicenedopanscheme}c   illustrate how  C substitutions to two silicene structures induce significant distortions caused by disorder; see SM for quantification of the disorder in the structures. In addition to the in-plane distortion, a large vertical bending is observed, indicating that the disorder caused by C doping is of long-range. In particular, position 1 exhibits the highest buckling distortion, which can extend afar from the substitution site since most of the distortions are elastic. Also, the farther away from the nanocluster's center the more stable the structure becomes, and the less disorder it presents. Furthermore, all of the doped configurations prefer to be in a triplet state indicating that these structures are ferromagnetic and stable under disorder. This intrinsic magnetism is an advantage since it is required for spin-based electronics;~\cite{han2014graphene,pesin2012spintronics} particularly information technology.~\cite{jansen2012silicon}

\begin{figure}[ht]
  \includegraphics[width=8.7cm]{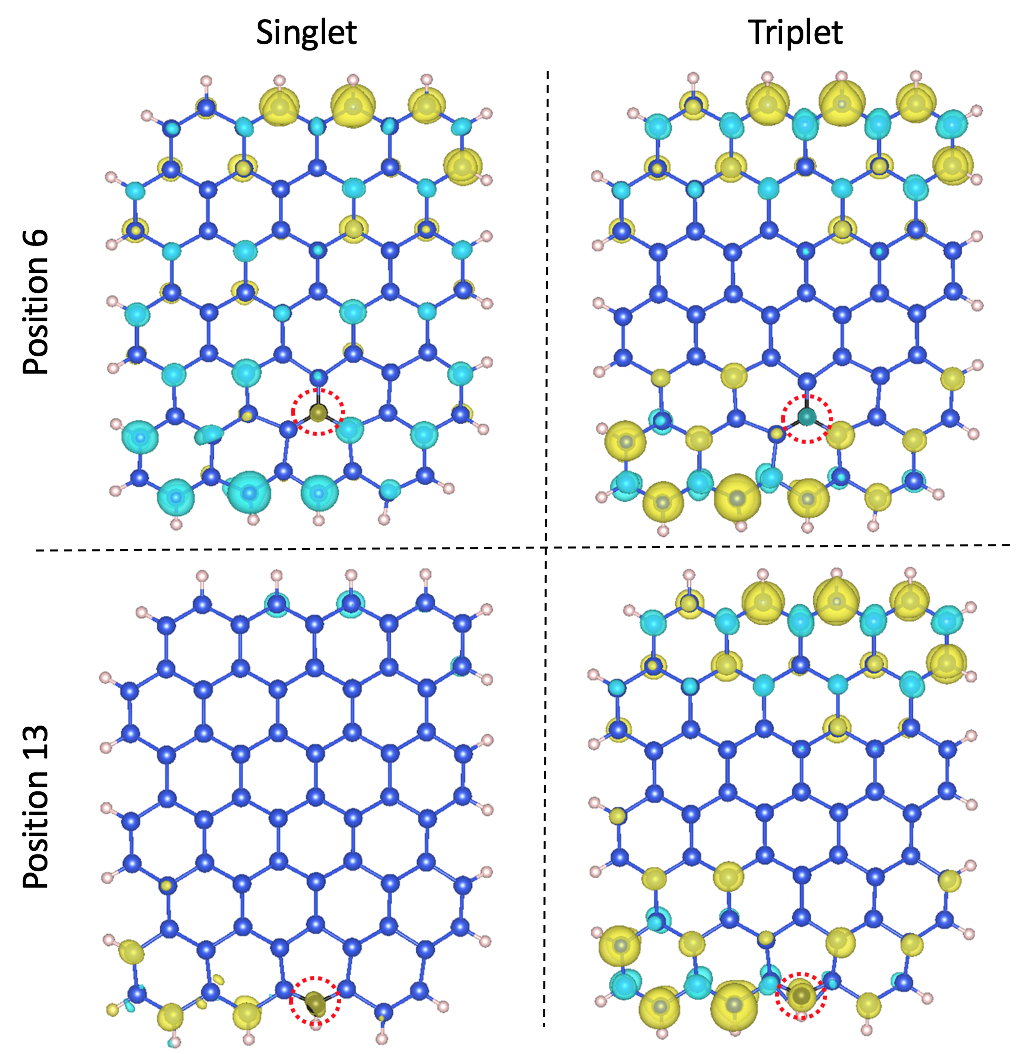}
  \caption{The spin density distributions for a single C-doped H-SiNC at two different site positions. The yellow (cyan) isosurface corresponds to the predominant spin-up (spin-down) charge density. The red circle highlights the position of the carbon atom in the silicene nanocluster.}
  \label{fig:spindensity}
\end{figure}

\begin{table}[ht]\small
\caption{Formation energies (in eV per impurity atom) of double C substitution at thirteen different sites of H-SiNC for singlet (S), triplet (T), and  quintuplet (Q) states at the B3LYP level of calculation. All formation energies were calculated with respect to the singlet undoped ground state of H-SiNC(4,6).}
\begin{tabular}{ccccccc}\\
    \toprule
Site & 2-C (S)  & 2-C (T) & 2-C (Q)  & 1+1 C(S)& 1+1 (T) & 1+1 C (Q) \\
\hline
 $1$ & $-5.710$ & $-5.849$ & $-4.992$ & $-5.638$ & $-5.793$ & -4.348 \\ 
 $2$ & $-5.752$ & $-5.890$ & $-5.097$ & $-5.703$ & $-5.840$ & -4.388 \\
 $3$ & $-6.177$ & $-6.350$ & $-5.525$ & $-5.830$ & $-5.957$ & -4.504\\
 $4$ & $-5.651$ & $-5.839$ & $-5.057$ & $-5.643$ & $-5.794$ & -4.337\\
 $5$ & $-5.655$ & $-5.831$ & $-5.001$ & $-5.666$ & $-5.803$ & -4.355\\
 $6$ & $-5.634$ & $-5.770$ & ---\footnotemark[1]     & $-5.699$ & $-5.870$ & -4.407\\
 $7$ & $-5.865$ & $-6.040$ & $-5.190$ & $-5.898$ & $-6.032$ & -4.580\\
 $8$ & $-5.997$ & $-6.147$ & $-5.418$ & $-5.840$ & $-5.979$ & -4.523\\
 $9$ & $-5.823$ & $-6.017$ & $-5.191$ & $-5.972$ & $-6.120$ & -4.646\\
 $10$& $-5.805$ & $-5.943$ & ---\footnotemark[1]      & $-5.904$ & $-6.053$ & -4.571\\
 $11$& ---\footnote{Structures did not converge.}      & $-5.952$ & $-5.155$ & $-5.846$ & $-5.986$ & -4.554\\
 $12$& Fixed & Fixed & Fixed & Fixed  & Fixed & Fixed\\
 $13$& $-6.592$ & $-6.690$ & $-5.828$ & $-6.605$ & $-6.725$ & -5.241\\
 $14$& $-6.242$ & $-6.360$ & $-5.550$   & $-6.221$ & $-6.492$ & -4.928\\
   \hline
    \end{tabular}
 \label{table2}
\end{table}

When a carbon atom replaces a silicon one, there is a redistribution of the electronic charge in the system. To illustrate this effect, we show the spatial spin density distribution in Fig.~\ref{fig:spindensity} for both singlet and triplet states at two different sites. For position  $S_6$, the singlet state exhibits a negative (positive) magnetic moment at the upper (lower) edge and a zero net magnetic moment indicating that at this position the C atom does not have a strong effect in the spin density compared with the pristine H-SiNC~\cite{pablo2018exploring}. In the triplet state, the nanoclusters now exhibit a net magnetic moment, having the spins paired-up at opposite edges. For the $S_{13}$ position in the singlet state, the spin density on the edge is locally concentrated. It is worth noting that such a spin density modification effect is also found in the other inner substitution configurations but it is gradually weakened as the C atom gets closer to the center of the nanocluster.

For two carbon substitutions, there will be new interactions that can lead to non-trivial effects in the electronic structure. To measure these interaction effects between two carbon atoms, one of them is fixed at position $S_{12}$ (the most stable position) while varying the other sites. The formation energies for double C-doped structures are listed in Table \ref{table2} for singlet and triplet states. All of these structures are thermodynamically favorable conforming that double C-doped H-SiNCs are stable. We also compare the stability of double versus single C-doping, by contrasting the energies $E_f (S_i+S_{12})$ with respect to $E_f(S_{12})+E_f(S_i)$ with $i=1,\ldots,14$. With the condition $|E_f (S_i+S_{12})|>|E_f(S_{12})+E_f(S_i)|$, only the $S_1$, $S_2$, $S_3$, $S_4$, and  $S_8$ positions can be stable for both singlet and triplet state configurations. Similarly to the single C-doping case, the formation energies increase as we move towards the center of the nanocluster.  For the adjacent doping cases, $S_{12}-S_{14}$, the $C-C$ bond introduces a higher strain into the structures, due to its shorter length, making those nanoclusters the most unstable ones. Additionally, we have included quintuplet states for the double C-doped nanoclusters since two triplet C defects can form a quintuplet (Q) state as well. Here, we can observe first that the formation energy of the quintuplet states follows a similar tendency as triplet and singlet states; second, quintuplet states are less stable than singlet and triplet states.

\begin{figure}[ht]
  \includegraphics[width=8.7cm]{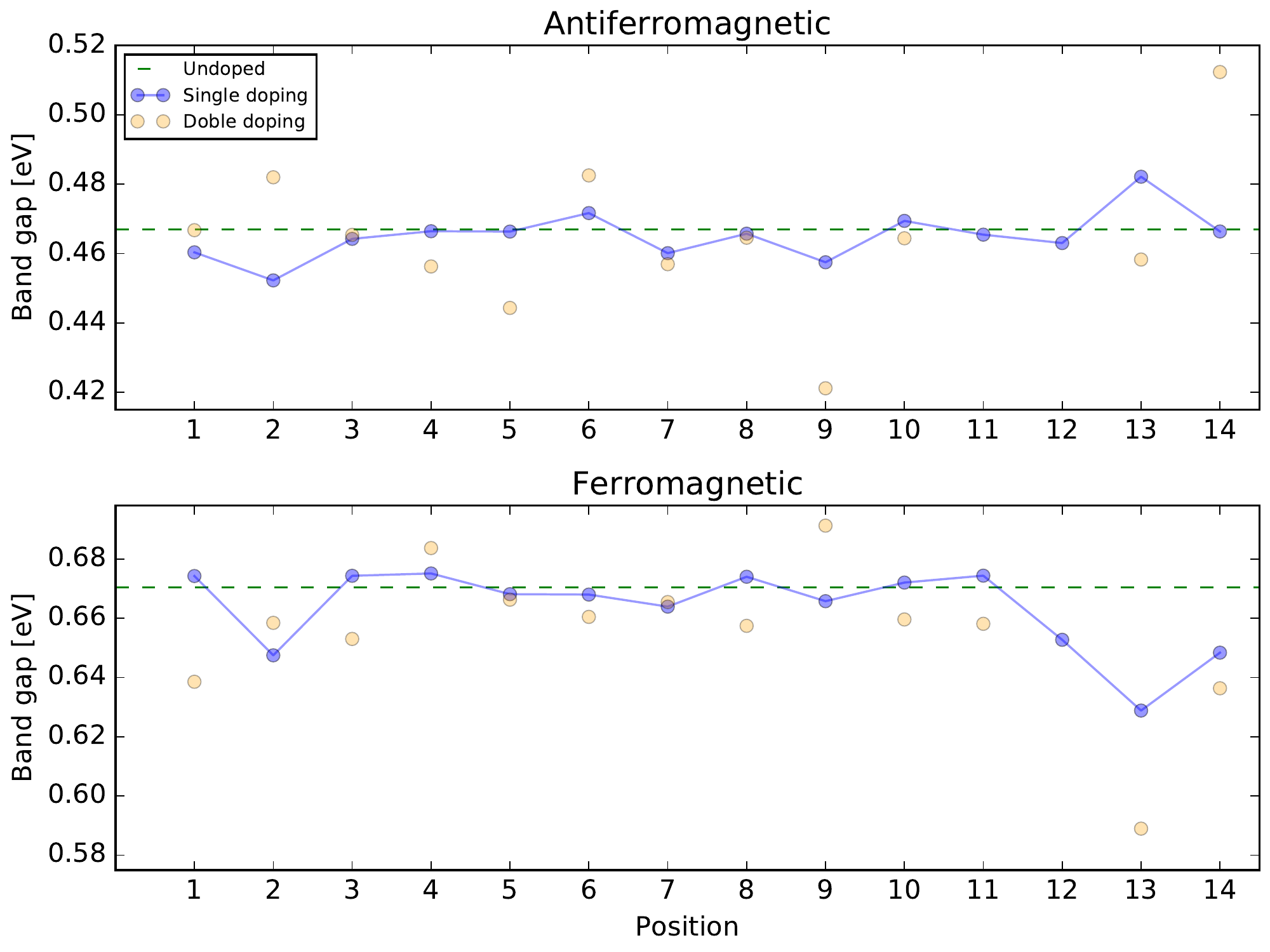}
  \caption{ {\bfseries Calculated band gap of a rectangular silicene nanocluster; H-SiNC for all fourteen  different positions.}  Antiferromagnetic and ferromagnetic states are illustrated  for single and doubled C-doped structures. The dashed green lines in both graphs represent the bang gap for the pristine cluster.}
  \label{fig:silicenebandgap}
\end{figure}

To show the effect of carbon interactions on the electronic structure of C-doped H-SiNCs, the HOMO-LUMO gap is calculated for all fourteen configurations, see Fig. \ref{fig:silicenebandgap}. 
For the antiferromagnetic states, the HOMO-LUMO gap depends on the site and the number of C atoms in the silicene nanocluster. Single C substitutions near the center have a larger effect on the band gap. For double C doping, the larger effects on the gap are shown near the edges due to correlation effects and structural relaxation. The results for the ferromagnetic states follow a similar trend to the antiferromagnetic state for single and double C substitutions.

\begin{figure}
  \includegraphics[width=8.7cm]{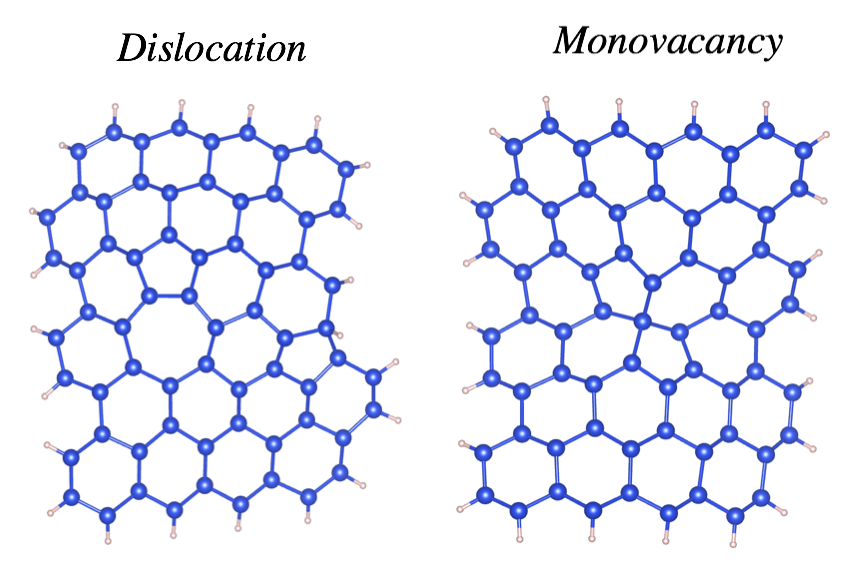}
  \caption{Geometry of a dislocation and a monovacancy in a silicene nanocluster. In both cases the structures shown are for singlet states configurations.}
  \label{fig:dislocation}
\end{figure}

Since dislocations and vacancies can arise as a consequence of imperfections during synthesis or stress imposed through thermal history,\cite{li2015defects} we also calculated the formation energy for a dislocation and a monovacancy at position $S_1$, see Fig.~\ref{fig:dislocation}. The formation energies for this dislocation are 5.082 and 5.021 eV, for the singlet and triplet state configurations respectively. The band gap for the singlet and triplet state are 0.562 and 0.558 eV respectively.
For a monovacancy, the formacion energy is 5.57 eV with a band gap of 0.405 eV. This shows the thermodynamically unstable nature of these defects. 

Nanoscale lengths structures could differ considerably from their bulk counterparts in their physical and electronic properties; hence, caution must be taken when extrapolating results from finite to periodic structures. With this in mind, we have studied periodic boundary conditions monolayers with supercell sizes of $2\times 2$,  $3 \times 3$, and  $4 \times 4$; see SM for more details. Formation energy and electronic structure properties were calculated using the PBE and B3LYP functionals. For the hybrid functional, B3LYP, we were able to obtain results for the $2\times 2$ supercell due to the high computational cost of these calculations for bigger systems. For PBE, we obtained the formation energies $-2.3609$ eV, $-2.4772$ eV, and $-2.5505$ eV for the $2\times 2$, $3\times 3$, and $4\times 4$ supercells respectively; thus, these structures are thermodynamically favorable, in agreement with the nanoclusters calculations. 
The B3LYP calculations are in qualitative agreement and differ only by small amount with respect to the PBE functional.

\begin{table}\small
\caption{Formation energies, $E_{f}$, and Band gaps, $E_{g}$, of double C substitution for spin-polarized systems at the PBE level of calculation. Here we use three different supercells. One C atom was fixed at position 1. The last row correspond to the monovacancy (MV) case.}
\begin{tabular}{l@{\hspace{0.5em}}cc@{\hspace{1em}}cc@{\hspace{1em}}cc}\\
    \toprule
 & \multicolumn{2}{c}{$2\times 2$} & \multicolumn{2}{c}{$3\times 3$} & \multicolumn{2}{c}{$4\times 4$} \\
\hline
 Site & $E_{f}$(eV) & $E_{g}$(eV) & $E_{f}$(eV) & $E_{g}$(eV) & $E_{f}$(eV) & $E_{g}$(eV) \\
 \hline
 $1$  & Fixed  & Fixed & Fixed & Fixed & Fixed & Fixed  \\
 $3$  & $-5.4917$ & 1.006 \footnote{Indirect band gap.} & $-5.0139$ & 0.142 & $-5.0616$ & 0.182  \\
 $4$  & $-4.0297$ & 0.199 \footnote{Direct band gap.} & $-4.1509$ & 0.041 & $-4.2305$ & 0.013  \\
 $6$  &    ---     &  ---       & $-4.8071$ & 0.050 & $-4.9504$ & 0.023  \\
 $7$  &    ---     &  ---       & $-5.6775$ & 0.000 & $-5.4399$ & 0.180  \\
 $8$  & $-6.2498$ & 0.001 \footnotemark[2] & $-5.6347$ & 0.050 & $-5.5299$ & 0.038  \\
 $9$  &    ---     &  ---       &    ---     &  ---   & $-4.9958$ & 0.187  \\
 $10$ &    ---     &  ---       &    ---     &  ---   & $-5.3934$ & 0.034  \\
 $11$ &    ---     &  ---       &    ---     &  ---   & $-5.4396$ & 0.183  \\
 MV   & $2.9956$  & 0.458 \footnotemark[2] &  $3.2189$ & 0.003 & $3.2918$  & 0.074  \\
   \hline
    \end{tabular}
 \label{blahperidoformation}
\end{table}

Formation energies for double C substitution are reported in Table~\ref{blahperidoformation}. Positions $S_{12}$ and $S_{13}$ are not considered here due to the size constraints of the supercells. B3LYP's results of double C substitution for a $2\times 2$ supercell are shown in SM. Formation energies shown in Table~\ref{blahperidoformation} follow the same trend as the nanoclusters. From the formation energies, we can see that the position $S_8$ is the most stable periodic structure as one may aspect from a doping configuration that preserves most of the global symmetry of the layer. Also, for those structures in which the carbon atoms are located at the same sublattice exhibit similar and larger band gaps than those for which the doping occurs in opposite sublattices. This symmetry-breaking doping induces a sublattice potential difference which in turn breaks the band degeneracy and opens the band gap.

\begin{figure*}[tbph!]
  \includegraphics[width=17cm]{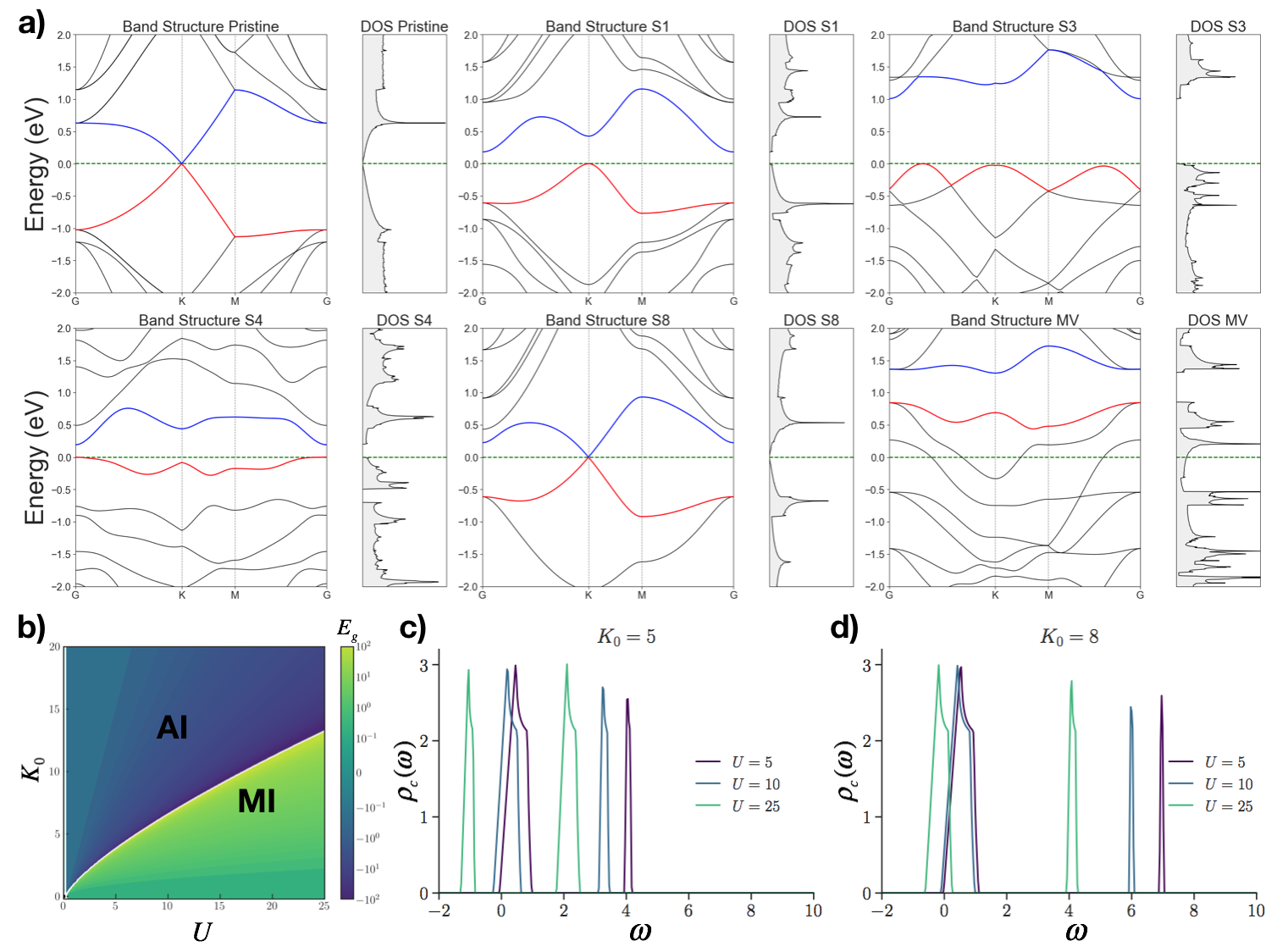}
  \caption{{\bfseries a)} The energy band structures and Density of States (DOS) of periodic monolayers with different doping positions. The last figure correspond to the Monovacancy (MV) case. The band  structure was calculated along the path of the high symmetry points $\Gamma-K-M-\Gamma$. The blue and red lines represent the conduction and valence  bands respectively. The zero in the energy axis is set at the Fermi level as shown by the dashed green line. {\bfseries b)} Phase diagram in the $K_0 U$ plane for the Anderson-Hubbard model on the honeycomb lattice. AI and MI refer to Anderson and Mott insulator, respectively. {\bfseries c)} LDOS of  C-doped silicene at $K_0=5$ for varying the Hubbard interaction $U$. {\bfseries d)} Same as {\bfseries c)} for $K_0=8$.}
  \label{fig:silicenebands}
\end{figure*}

\textit{Correlating disorder and interactions with electronic band structures.}  As mentioned earlier, carbon substitutions introduce disorder and interactions that can affect the electronic structure. Here we address the question of how the details of the band structure near the Dirac point is affected by disorder and interactions. 

We select only  the $2\times 2$ supercell using the PBE functional.  Figure~\ref{fig:silicenebands}a) shows the bandstructure of different silicene structures alongside their respective density of states (DOS). In the following discussion, $S_1$ represents the only single doped structure while the rest of the mentioned positions stand for a double doped structure with position $S_1$ fixed. Note that in the bandstructure calculations, different carbon positions generate different disorder and interaction strength which introudce a shift of the energy bands respect to a reference case, i.e. pristine silicene.  For instance, when one silicene atom in the supercell is substituted by a carbon atom, a gap opens at the K point  of 0.181 eV with a semiconductor behavior. Depending on the sublattice position, adding an additional carbon atom could break the symmetry of the system, even more, resulting and large band gap; this is the case for the $S_1-S_3$ system (same sublattice) which shows an indirect gap of 1.006 eV and semiconductor behavior. On the other hand, the systems $S_1-S_4$ and $S_1-S_8$ act on different sublattices and show band gaps of 199 and 1 meV respectively. Both systems present a direct gap; yet, the C4 gap is around the G point and C8 in the K direction. For the MV structure, we observe  a metallic behavior, with a sizeable direct band gap of 0.458 eV. Furthermore, we do not observe self-healing for any of the MV systems. 

\textit{Understanding the phase transition}. Using a toy model that captures the strength of electron interaction and the amount of disorder we have calculated the local density of states (LDOS) around the Dirac point for the C-doped silicene structure. The LDOS in this model is

\begin{equation}
    \rho_c(\omega)=\int \frac{d^2k}{(2\pi)^2}\delta \left(\omega-\epsilon_\mathbf{k}-E_g(\omega,K_0,U)\right)
    \label{spectral}
\end{equation}
With 
\begin{equation}
E_g(\omega,K_0,U)=\frac{K_0^2}{\omega-\tilde{\epsilon_p}+\frac{2U}{(\hbar v_F)^2}\frac{K_0^2}{(\omega-\tilde{\epsilon_p})^2}\left(\omega-\frac{K_0^2}{\omega-\tilde{\epsilon_p}}\right)}. 
\label{gap}
\end{equation}

where $\tilde{\epsilon_p}=\epsilon_p-\frac{U}{2}$,  $\epsilon_\mathbf{k}=\hbar v_F |\mathbf{k}|$ is the energy dispersion at the $K$-point and $K_0^2=\langle U_{imp}(\mathbf{r}_i)U_{imp}(\mathbf{r}_j)\rangle$ is the disorder correlator with $U_{imp}(\mathbf{r}_i)$ being the disorder potential strength. The energy of the $p$-electron is given by $\epsilon_p$. The Hubbard $U$ parameter describes the on-site interaction between two electrons with opposite spins on site $i$. 

Using the proposed model we obtained the phase transition diagram for the Hubbard interaction $U$ vs the strength of the disorder $K_0$ at zero temperature, see Fig.~\ref{fig:silicenebands}b). Here, the two insulating phases are separated roughly by $U \propto 2 K_0$. This means that the disorder affects the Mott transition by pushing it to larger interaction strength as also reported in refs. \cite{PhysRevLett.94.056404} and \cite{PhysRevB.95.045130}. Figures ~\ref{fig:silicenebands}c) and \ref{fig:silicenebands}d) show the double-peak impurity resonances around the Fermi level for different disorder and interaction strengths. This type of double-peak structure has been experimentally observed in nitrogen-doped graphene, and most likely originates from long-range interactions.~\cite{PhysRevB.85.161408} Indeed, the peak-peak separation goes down when increasing $U$. Eventually, for large enough $U$, one of the resonance peaks moves past $\omega=0$, where the Dirac point of the unperturbed system is located. In addition, Fig. \ref{fig:silicenebands}d) reveals that there are more electronic states close to the Fermi level for higher disorders. This alternation between lower and higher densities of states brings out the fact that the two sublattices of silicene are affected differently by a substitutional atom (see SM for consulting $K_0$ values). Hence, both the magnitude and the spatial range of the impurity potential are modified by interactions. Those results alongside DFT calculations, help us to understand the driven factor for each system's transition.

\textbf{In summary}, we study the effects of disorder on the electronic properties of silicene nanoclusters and monolayers caused by C doping. Total energies analysis indicates that C tends to be doped at the edges of finite silicene structures, which are ferromagnetic. Using field theory, we are able to describe the driven factor for different dope-induced transitions and to contrast its predictions with DFT results. Specifically, by looking at the band gap of the systems we are able to distinguish between Anderson and Mott transitions. For a pristine silicene monolayer, the electronic dispersion is characterized by Dirac cones. Systems were the Hubbard interaction predominates break the degeneracy and crossing of the valence and conduction bands, hence opening a band gap; however, as the strength of the disorder increases, the gaps close eventually recovering the presence of Dirac cones. This indicates that the Mott and Anderson insulators are continuously connected. The framework of our analysis not only apply to silicene but also other doped 2D materials.~\cite{usachov2011nitrogen} Moreover, this combination of theoretical and computational results may be valuable in the design of Silicene-based electronic devices in spintronics.~\cite{han2014graphene,pesin2012spintronics}



\begin{acknowledgments}
R.P.-P, J. K., and T.V.V. acknowledge the King Abdullah University of Science and Technology for support under contract (OSR- 2015-CRG4-2634). R.P.-P thanks the support from FEMSA and ITESM. J.L.M.-C. acknowledges start-up funds from Florida State University (FSU) and the Energy and Materials Initiative and facilities at the High Performance Material Institute(HPMI). Some of the computing for this project was performed on the HPC cluster at the Research Computing Center at the Florida State University (FSU).  A portion of this work was performed at the National High Magnetic Field Laboratory, which is supported by National Science Foundation Cooperative Agreement No. DMR-1644779 and the State of Florida. 
\end{acknowledgments}

\bibliography{references}

\pagebreak
\widetext
\begin{center}
\textbf{\large Supplementary Material for "Understanding disorder in 2D materials: the case of carbon doping of Silicene"}
\end{center}
\setcounter{section}{0}
\setcounter{equation}{0}
\setcounter{figure}{0}
\setcounter{table}{0}

\makeatletter
\renewcommand{\theequation}{S\arabic{equation}}
\renewcommand{\thefigure}{S\arabic{figure}}
\renewcommand{\thetable}{S\arabic{table}}
\renewcommand{\bibnumfmt}[1]{[S#1]}
\renewcommand{\citenumfont}[1]{S#1}
\renewcommand{\angstrom}{\textup{\AA}}

\section{Electronic properties of different shapes}

\begin{figure}[ht]
  \includegraphics[width=12cm]{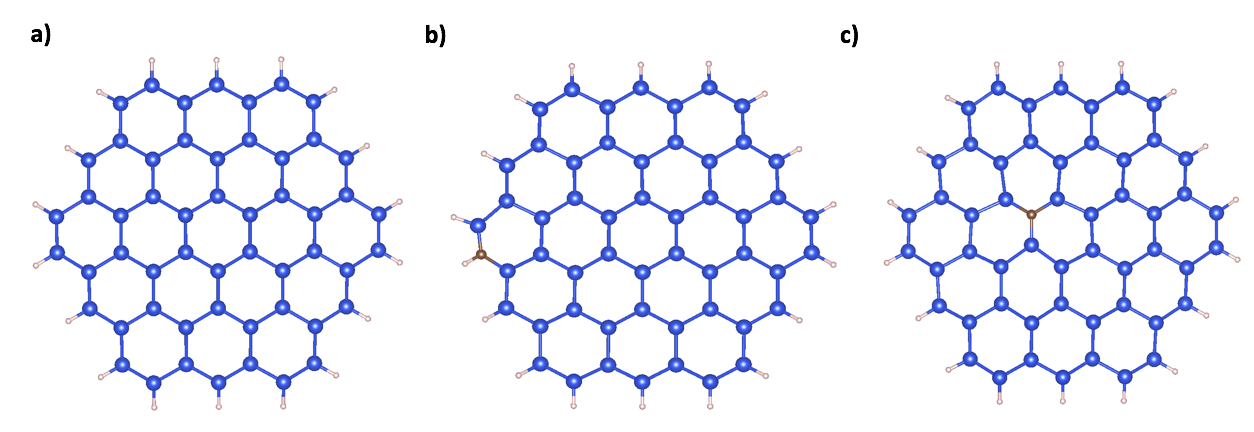}
  \caption{Geometric structures of pure and doped hexagonal silicene nanoclusters.   Coloring scheme: H; white, Si; blue, C; brown.}
  \label{fig:hexagonal}
\end{figure}

\begin{figure}[ht]
  \includegraphics[width=12cm]{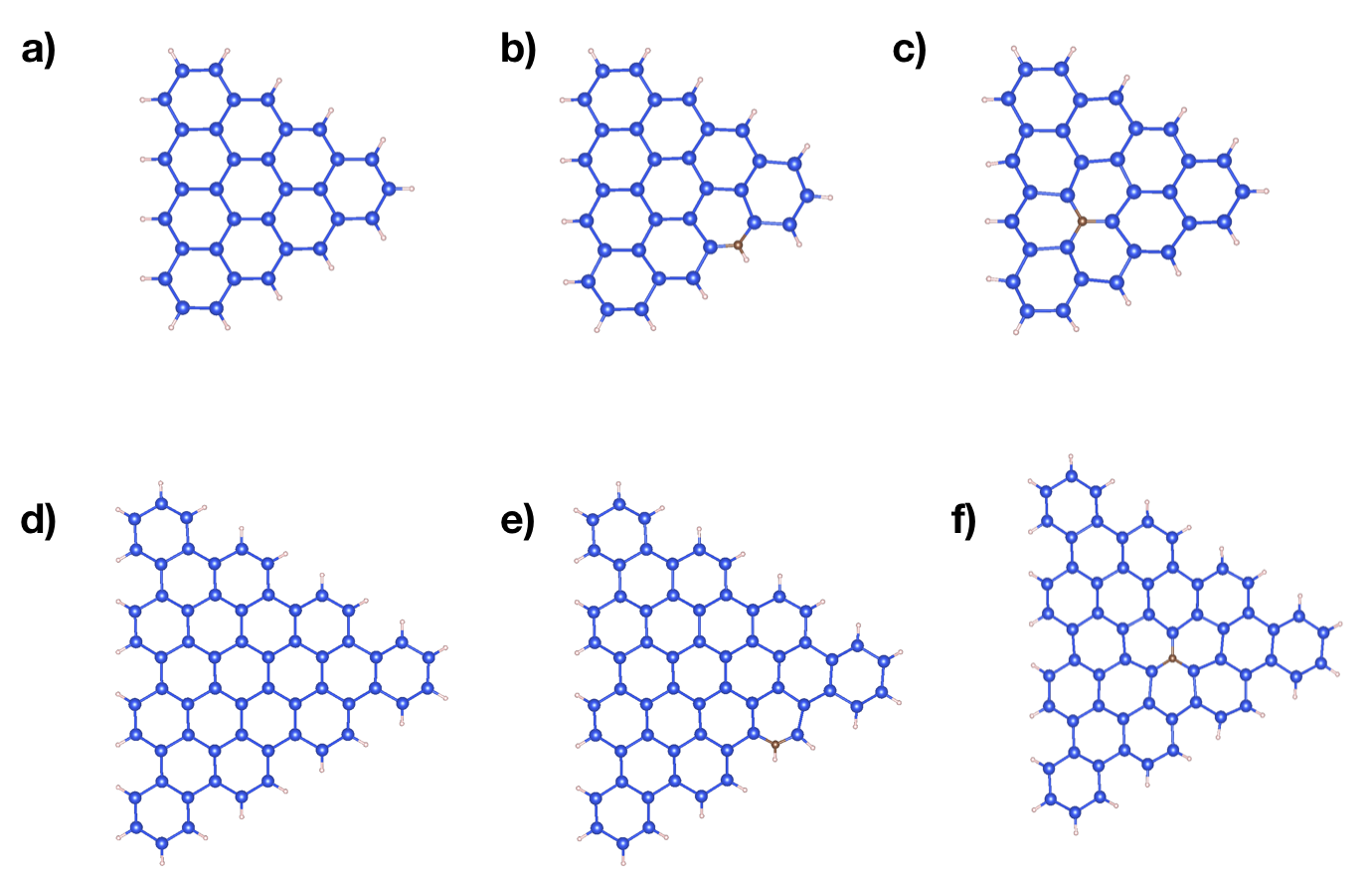}
  \caption{Geometric structures of pure and doped triangular silicene nanoclusters.   Coloring scheme: H; white, Si; blue, C; brown.}
  \label{fig:Triangular}
\end{figure}

To demonstrate the effect of doping on the stability of different shapes, we calculate the formation energies for different geometries which are shown in Figs.~\ref{fig:hexagonal} and \ref{fig:Triangular}.  

 According to table~\ref{table112}, doped structures with a hexagonal shape are more stable in the singlet state. These results are in agreement with previous theoretical findings and the Liebs' theorem for a bipartite lattice$^{1,2}$.

\begin{table}[ht]
\caption{ The formation energies (in eV) for the hexagonal structures shown in Fig.~\ref{fig:hexagonal} for pure and single doped structures.  All the formation energies were calculated with respect to the pure structure in the singlet state.}
\begin{tabular}{ccc}\\
    \toprule
Figure & $E_f$ (S=0)  & $E_f$(S=1)  \\
\hline
\ref{fig:hexagonal}a) & $0.000$ & $0.720$  \\ 
 \ref{fig:hexagonal}b) & $-3.330$ & $-2.601$ \\
\ref{fig:hexagonal}c) & $-2.272$ & $-1.581$  \\
   \hline
    \end{tabular}
 \label{table112}
\end{table}

For triangular structures (Fig.~\ref{fig:Triangular}), the multiplicity differs from hexagonal and rectangular shapes since triangular structures can have a different number of Si atoms in each sublattice ($N_A \neq N_B$), therefore, the multiplicity and the total spin moment vary. For instance, the structure shown in Fig.~\ref{fig:Triangular} can have two possible multiplicities $S'=2S+1=2$  and  $S'=4$ with $S=(N_A-N_B)/2=3/2$, $N_A=18$ and $N_B=15$. The corresponding multiplicities for the structure shown in Fig.~\ref{fig:Triangular}d) are $S=0$  and $S'=1$ since $N_A=N_B=30$.  From table \ref{table223}, we can observe that multiplicity stability vary depending on the number of Si atoms in each sublattice. These ground states are in good agreement with Lieb's theorem$^2$.

\begin{table}[ht]
\caption{ The formation energies (in eV) for  two different triangular structures shown in Fig. 2 for pure and doped structures. Formation energies were calculated with respect to the pure structure.}
\begin{tabular}{cccccc}\\
    \toprule
Figure & $E_f$ (S=2)  & $E_f$(S=4) & Figure & $E_f$ (S=1) & $E_F$ (S=3) \\
\hline
\ref{fig:Triangular}a) & $0.000$ & $-0.224$ &\ref{fig:Triangular}d) & $0.000$ & $0.921$   \\ 
 \ref{fig:Triangular}b) & $-3.215$ & $-3.409$ &\ref{fig:Triangular}e) & $3.344$ & $-2.442$  \\
\ref{fig:Triangular}c) & $-2.322$ & $-2.527$  &\ref{fig:Triangular}f) & $-2.289$ & $-1.435$  \\
   \hline
    \end{tabular}
 \label{table223}
\end{table}

\section{Computational details for Periodic silicene structures}

\begin{figure}[ht]
  \includegraphics[width=12cm]{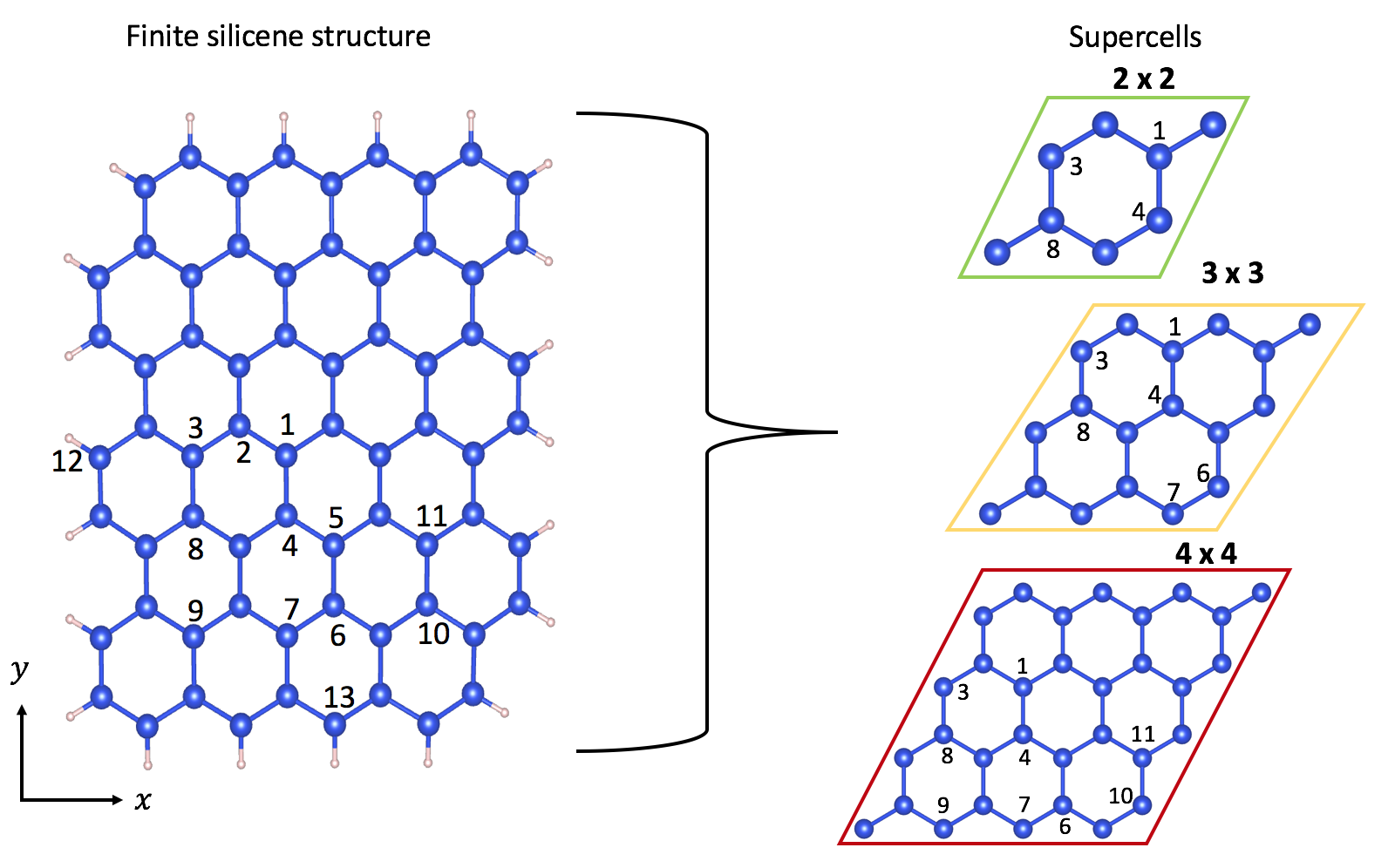}
  \caption{The three different supercells taken from the silicene nanocluster.}
  \label{fig:supercells}
\end{figure}

For the periodic structures, we considered three different supercells $2 \times 2$,  $3 \times 3$, and  $4 \times 4$, see Fig.~\ref{fig:supercells}. The Brillouin zone was sampled by a 11x11x1 $\Gamma$ centered k-points mesh in the Monkhorst-Pack scheme. Structures are optimized by minimizing the forces on individual atoms below 0.001 eV\,\AA$^{-1}$, and the convergence self-consistent energy of every electron step is less than $10^{-5}$ eV using PBE and B3LYP functionals. For the PBE functional, the generalized gradient approximation (GGA) is carried out. The projector-augmented wave (PAW) method is used to describe the electron-ion potential. The semi-empirical Grimme-D3 dispersion corrections were also added in the present calculations to incorporate van der Waals dispersion effects on the system, within the default VASP parameters. A Gaussian smearing of $0.05$ eV for geometry optimization was taken. The $3 s^2 3 p^2$, $3 s^2 3 p^2$, and $1s^2$ electrons are taken as the valence electrons of Si, C, and H atoms, respectively. To avoid layer-layer interaction a vacuum of 10 A above and below the layers was added. In addition, the projector-augmented wave (PAW) method is used to describe the electron-ion potential$^3$.

\section{Results of B3LYP functional for periodic structures}
\label{D1}

To prove that our conclusions are not entirely dependent on the type of functional selected for the periodic structures, we compare the results of PBE with the B3YLP functional.  This hybrid functional was chosen because it gives encouraging results for the average calculated nearest Si-Si distance results as compared to a silicene nanoribbon on Ag (110) ($2.24$ \AA). 
Then, we calculate the formation energies for a $2\times 2$ supercell structure to show that the tendency of formation energies are not dependent on the functional used for its calculation.  In table \ref{blahperidoformationB3LYP}, we show the formation energies of a double C  doped structure using B3LYP and PBE functionals. For the doped systems, B3LYP adds a small correction to energy, while for the monovacancy the correction is much greater. Overall, the qualitative trend from PBE is conserved.  

\begin{table}[ht]
\caption{Formation energies, $E_{f}$, of double C substitution for spin-polarized systems at the B3LYP and PBE  level of calculation. Here we use a $2\times 2$ supercell. One C atom was fixed at position 1. The last row correspond to the monovanacy (MV) case.}
\begin{tabular}{l@{\hspace{0.5em}}c@{\hspace{1em}}c@{\hspace{1em}}}\\
    \toprule
 & B3LYP &  PBE \\
\hline 
 Site & $E_{f}$(eV) &$E_{f}$(eV) \\
 \hline
 $1$  & Fixed  & Fixed  \\
 $3$  & $-5.9061$  & $-5.4917$ \\
 $4$  & $-4.0867$  & $-4.0297$ \\
 $6$  &    ---   &  ---   \\
 $7$  &    ---   & ---   \\
 $8$  & $-6.7483$ & $-6.2498$  \\
 $9$  &    ---    &   --- \\
 $10$ &    ---   &  --- \\
 $11$ &    ---   &  ---  \\
 MV   & $10.2520$ & $2.9956$  \\
   \hline
    \end{tabular}
 \label{blahperidoformationB3LYP}
\end{table}

\section{Strain effects in C-doped Silicene structures}
The substitution of Si atoms by C adds uniaxial strain effects on the structures. To quantify the extent of deformation on the structures, we have calculated the average strain, $\overline{\epsilon}$, as follows  

\begin{equation}
    \overline{\epsilon}= \frac{1}{3}\sum_{i=1}^3\frac{a_i-a_i^0}{a_i^0}=\sum_{i=1}^3 \epsilon_i
\end{equation}

\begin{table}\small
\caption{The average strain effect for all thirteen positions for a single C substitution. The singlet and triplet configurations are represented by $S=0$ and $S=1$, respectively.}
\begin{tabular}{ccc}\\
    \toprule
Site &  $\overline{\epsilon}$ (S=0)& $\overline{\epsilon}$ (S=1) \\
\hline
 $1$ &  $-0.1753$ & $-0.1758$\\ 
 $2$ & $-0.175$ & $-0.175$\\
 $3$ & $-0.1820$ & $-0.182$\\
 $4$ & $-0.1761$ & $-0.176$\\
 $5$ & $-0.1760$ & $-0.175$\\
 $6$ & $-0.1765$ & $-0.176$\\
 $7$ & $-0.1780$ & $-0.1785$\\
 $8$ & $-0.1790$ & $-0.1784$\\
 $9$ & $-0.1840$ & $-0.183$\\
 $10$& $-0.1810$ & $-0.181$\\
 $11$ & $-0.1780$ & $-0.178$\\
 $12$ & $-0.2230$ & $-0.223$\\ 
 $13$ & $-0.1990$ & $-0.200$\\
  $14$ & $-0.1970$ & $-0.1967$\\
   \hline
    \end{tabular}
 \label{table3}
\end{table}

\noindent where $a_i^0$ ($a_i$) is the equilibrium (strained) bond length of each nearest neighbor atom, $i$, for each carbon atom in the doped structure. 
This consideration is used because the larger lattice distortions are seen at the vicinity of the C substitution, see Figs.~\ref{fig:silicenedopanscheme}b) and  \ref{fig:silicenedopanscheme}c). As can be seen in Table \ref{table3}, the average strain follows a similar trend to the energy. The higher strain the lower the energy in the nanocluster. It seems that to accommodate the compressive epitaxial strain, a relatively large buckling is induced while keeping the Si bond lengths constant.

Additionally, we compute the strain effect for  double C-doped periodic structures and the MV structure, which are shown in Table~\ref{strainperidic}. The results show that a higher strain leads to lower energy in the C-doped periodic structures in most of the cases. These results are in agreement with the doped finite structure's results. For the monovacancy's structure, the strain is positive because the Si bond lengths increase contrary to the doped structures.

\begin{table}\small
\caption{The strain effect on double C substitution of the periodic structures.}
\begin{tabular}{cccc}\\
    \toprule
Site &   $\overline{\epsilon}$ ($2\times 2$) &$\overline{\epsilon}$ ( $3\times 3$) & $\overline{\epsilon}$ ($4\times 4$)\\
\hline
 $1$ &  Fixed & Fixed & Fixed \\ 
 $3$ & $-0.162$ & $-0.175$ & $-0.178$\\
 $4$ & $-0.180$ & $-0.196$ & $-0.199$\\
 $6$ & ---      & $-0.176$ & $-0.178$\\
 $7$ & ---      & $-0.190$ & $-0.188$\\
 $8$ & $-0.204$ & $-0.190$ & $-0.189$\\
 $9$ & ---      & ---      & $-0.181$\\
 $10$& ---      & ---      & $-0.188$\\
 $11$& ---      & ---      & $-0.188$\\
 MV  &  $0.023$ & $0.026$  & $0.035$ \\
   \hline
    \end{tabular}
 \label{strainperidic}
\end{table}

\section{Understanding phase boundary: a Toy model for weak interaction}

\begin{figure}[ht]
  \includegraphics[width=12cm]{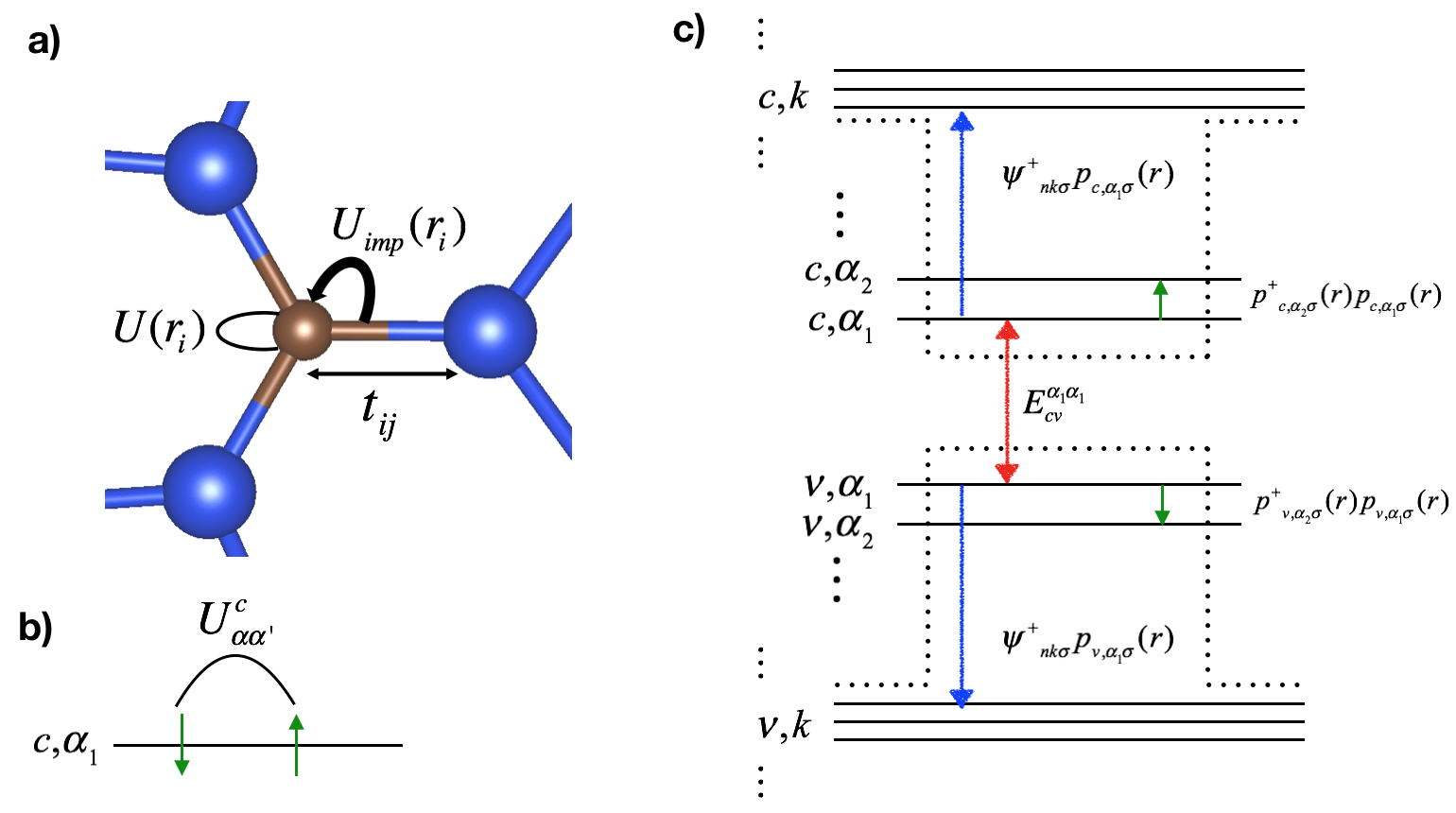}
  \caption{a) Scheme of silicene with an impurity atom. Here, $t_{ij}$, $U_{imp}(\mathbf{r}_i)$, and $U(\mathbf{r}_i)$ are parameters shown in Eq.~\ref{Hamiltonian}. b) Level scheme of the impurity-continuum model:  the hopping paramets have a crystal momentum $\mathbf{k}$. $\alpha$ represents the $p$-energy levels. The band gap between conduction and valence band states is denoted by $E_{cv}^{\alpha_1 \alpha_1}$. c) The electron-electron interaction on the $p$-energy level with magnitud $U_{\alpha \alpha' \alpha' \alpha}^c$.}
  \label{fig:stheory}
\end{figure}

Our model is schematically shown in Fig. \ref{fig:stheory}a). We consider fermions in a two-dimensional lattice with $i$ being the site index. Within each site, there is spin or pseudo-spin (such as two sublattices within each unit cell) degree of freedom denoted by $\sigma=\uparrow,\downarrow$. We consider that the hopping only takes place between the next neighbor atoms; thus, the single-particle Hamiltonian is written as 

\begin{equation}
    \hat{h}=-\sum_{\langle i,j\rangle}\sum_{\sigma}t_{ij}^\mu \psi_{\mu,\sigma}^\dagger (\mathbf{r}_i) \psi_{\mu,\sigma}(\mathbf{r}_j)
\end{equation}

Here, $t_{ij}^\mu$ is the hopping parameter with band index $\mu \in \{c,v\}$ and $\psi_{\mu,\sigma}^\dagger(\mathbf{r}_i) (\psi_{\mu,\sigma}(\mathbf{r}_i))$ is the creation (anhilation) fermionic operator for an electron on site $i$ with spin $\sigma$ and band index $\mu$ $^4$.

Now, to study the effect of a carbon atom in the silicene structure, we first introduce an on-site $p$-electron energy and on-site electron-electron interaction, that is 

\begin{equation}
    \hat{H}_{p}=\epsilon_p^\mu \sum_{i=1}^{N}\sum_{\sigma} p_{\mu,\sigma}^\dagger (\mathbf{r}_i) p_{\mu,\sigma}(\mathbf{r}_i)+\sum_{i=1}^N \sum_{\sigma \sigma' } \frac{U_{\sigma \sigma' \beta}^\mu(\mathbf{r}_i)}{2}p_{\mu,\sigma}^\dagger (\mathbf{r}_i)p_{\mu,\sigma'}^\dagger(\mathbf{r}_i) p_{\mu,\sigma'}(\mathbf{r}_i)p_{\mu,\sigma}(\mathbf{r}_i)
\end{equation}

where $p_{\mu,\sigma}^\dagger (\mathbf{r}_i)$ and $p_{\mu,\sigma}(\mathbf{r}_i)$ are the creation and anhilation operators for an electron  in orbital $p$ on site $i$  with spin $\sigma$ and band index $\mu$; $\epsilon_p^\mu$ is the p-electron energy level with the band index $\mu$.  Therefore, $H_{p}$ describes the atomic limit of an isolated atom involving a Kramers doublet of  energy $\epsilon_p$. 
If $U_{\sigma \sigma'}^\mu(r_i)=U$ is constant, we can  obtain the Hubbard model with local disorder (Anderson-Hubbard model) $^5$. Figure \ref{fig:stheory}b) illustrates the electron-electron interaction on the conduction band. 

When the carbon atom substitutes a silicon atom in the silicene structure (sea of electrons), the $p$-electrons within the core of the atom can tunnel out, hybridizing with the Bloch states of the surrounding medium, see Fig. \ref{fig:stheory}a). Therefore, we  include a hybridization term between conduction (valence) electron and the local $p$-electron caused by a carbon impurity as follows 

\begin{equation}
    \hat{H}_{hyb}=\sum_{i=1}^N\sum_{\sigma}
 U_{imp}(\mathbf{r}_i) \left[\psi_{\mu,\sigma}^\dagger (\mathbf{r}_i) p_{\mu,\sigma}(\mathbf{r}_i)+ p_{\mu,\sigma}^\dagger (\mathbf{r}_i) \psi_{\mu,\sigma}(\mathbf{r}_i)\right].
\end{equation}
 
This term is the result of applying first-order perturbation to the degenerate states of the conduction (valence) band and the atomic $p$-orbital.  If we choose  the impurity's position  to be random, we can select a  randomly fluctuating $U_{imp}(\mathbf{r}_i)$ of the form$^6$: 

\begin{equation}
    U_{imp}(\mathbf{r}_i)=\sum_{k=1}^{N_{imp}}U_ke^{-|\mathbf{r}_i-\mathbf{r}_k|^2/2\zeta}
\end{equation}

where $N_{imp}$ is the number of impurities, $U_k$ is the potential amplitude at the $k$-th site in the interval $(-\delta,\delta)$, and $\zeta>> \sqrt{3}a$ is the correlation length with $a$ being the lattice constant. In the special case $\zeta << a$, $N_{imp}=N$ each of the lattice sites in the  structure has a randomly fluctuating potential. We quantify the disorder strength by the dimensionless correlator $^7$ 

\begin{equation}
    K_0^2= \frac{V}{(\hbar v_F)^2} \frac{1}{N^2} \sum_{i=1}^N \sum_{j=1}^N \langle U_{imp}(\mathbf{r}_i) U_{imp}(\mathbf{r}_j) \rangle = \frac{1}{9}\sqrt{3}(\delta/\tau)^2 \left( \frac{1}{NN_{imp}}\right)\left(  \sum_{n=1}^{N_{imp}} \sum_{i=1}^N \text{exp} \left( -\frac{|\mathbf{r}_i-\mathbf{R}_n|^2}{2\zeta^2}\right) \right)^2
    \end{equation}

of the random impurity potential (with vinishing mean-value, $\langle U_{imp}(\mathbf{r}_i)\rangle=0$). Here, we normalize the correlator with respect to the Fermi velocity. 

The total Hamiltonian is therefore given by 

\begin{equation}
    \hat{H}=\hat{h}+\hat{H}_p+\hat{H}_{hyb}. 
\end{equation}

Here we have set the chemical potential $\mu=0$. This is crutial to ensure that the ground state with $\hat{h}$ alone is a semimetal. Figure \ref{fig:stheory}c) show the total Hamiltonian with all of its interactions.  

Before proceeding, it should be pointed out that we only perform analytical calculations for the conduction band, (i.e. $\psi^\dagger_{c,\sigma}\equiv \psi_{\sigma}$, $\psi_{c,\sigma}\equiv \psi_{\sigma}$ $p^\dagger_{c,\sigma}\equiv p_{\sigma}$, and $p_{c,\sigma}\equiv p_{\sigma}$). In addition, the two-point Green's function  are diagonal  because the disorder average over the $U(\mathbf{r}_i)$ above forces the incoming and outgoing time frames to be identical. 

We then turn to the path integral formalism of our model, whose partition function is given by 

\begin{equation}
\begin{split}
    Z&=\int D\psi^\dagger D\psi Dp^\dagger Dp e^{-S_0-S_{imp}-S_U}\\
    S_0&= \int_0^\beta d\tau \left[ \sum_{ij}\sum_{\sigma} \psi_{\sigma}^\dagger(\mathbf{r}_i,\tau) \left(\partial_\tau \delta_{ij}-t_{ij}\right)\psi_{\sigma}(\mathbf{r}_j,\tau)+\sum_j\sum_\sigma p_{\sigma}^\dagger(\mathbf{r}_i,\tau)(\partial_\tau+\epsilon_p)p_\sigma(\mathbf{r}_i,\tau) \right]\\ 
    S_{imp}&=\int_0^\beta d\tau \sum_i \sum_\sigma U_{imp}(\mathbf{r}_i) \left[ \psi_{\sigma}^\dagger(\mathbf{r}_i,\tau)p_\sigma(\mathbf{r}_i,\tau) +p^\dagger_\sigma(\mathbf{r}_i,\tau)\psi_{\sigma}(\mathbf{r}_i,\tau)   \right]\\
    S_{U}&= \int_0^\beta d\tau \sum_i \sum_{\sigma \sigma'}\frac{U_{\sigma \sigma'}(\mathbf{r}_i)}{2}p_{\sigma}^\dagger (\mathbf{r}_i,\tau)p_{\sigma'}^\dagger(\mathbf{r}_i,\tau) p_{\sigma'}(\mathbf{r}_i,\tau)p_{\sigma}(\mathbf{r}_i,\tau). 
    \end{split}
\end{equation}

After performing the standard Gaussian random average over each independent $U_{imp}(\mathbf{r}_i)$  and focusing on one replica realization,$^8$  we obtain

\begin{equation}
    \begin{split}
        Z&=\int D\psi^\dagger  D\psi Dp^\dagger Dp e^{-S_0-S_{imp}-S_{U}}\\
        S_{imp}&=-K_0^2\int_0^\beta d\tau \int_0^\beta d\tau' \sum_i \sum_\sigma \psi_{\sigma}^\dagger(\mathbf{r}_i,\tau) p_{\sigma}(\mathbf{r}_i,\tau) p_{\sigma}^\dagger(\mathbf{r}_i,\tau') \psi_{\sigma}(\mathbf{r}_i,\tau')\\
    \end{split}
\end{equation}

To decopled the $S_U$ term, we can rewrite the Hubbard interaction in terms of the spin operator$^9$

\begin{equation}
n_{ \downarrow}\left(\mathbf{r}_{i}\right) n_{ \uparrow}\left(\mathbf{r}_{i}\right)=+\frac{1}{2}\left(n_{ \downarrow}\left(\mathbf{r}_{i}\right)+n_{ \uparrow}\left(\mathbf{r}_{i}\right)\right)^{2}-\frac{1}{2}\left(n_{ \downarrow}\left(\mathbf{r}_{i}\right)+n_{ \uparrow}\left(\mathbf{r}_{i}\right)\right)
\end{equation}

where we have used the fact that $n_{ \uparrow}^2\left(\mathbf{r}_{i}\right)=n_{ \uparrow}\left(\mathbf{r}_{i}\right)$  with $n_{\sigma}(\mathbf{r}_i)=p_{\sigma}^\dagger(\mathbf{r}_i)p_{\sigma}(\mathbf{r}_i)$. Then we use the Hubbard–Stratonovich (HS) transformation, 

\begin{equation}
\begin{split}
    e^{-\sum_{i,\sigma} \frac{U}{2} n_{\sigma}(\mathbf{r}_i,\tau)n_{-\sigma}(\mathbf{r}_i,\tau)}&= \int D\phi \text{exp} \left(-\frac{1}{2U}\sum_{i\sigma}\phi_{\sigma}(\mathbf{r}_i,\tau) \phi_{-\sigma}(\mathbf{r}_i,\tau) +\sum_{i\sigma}i\phi_{\sigma}(\mathbf{r}_i,\tau) n_{-\sigma}(\mathbf{r}_i,\tau) \right)
    \end{split}
\end{equation}

To obtain the effective partition function, we integrate out all fermions leading to 

\begin{equation}
    \begin{split}
        Z=\int D \phi  e^{-S_{eff}}\\
        S_{eff}=-\sum_{ij\sigma}\text{Tr}\ln \left[(\partial_\tau-t_{ij}) \left(\partial_\tau \delta_{ij}+\epsilon_p \delta_{ij}-i\phi_{-\sigma}(\tau)\delta_{ij}-\frac{U}{2}\delta_{ij}\right)-K_0^2\delta_{ij}\right] \\
   +\frac{1}{2U}\int_0^\beta d\tau \sum_{i\sigma}\phi_{\sigma}(\mathbf{r}_i,\tau) \phi_{-\sigma}(\mathbf{r}_i,\tau)
    \end{split}
\end{equation}

where $S_{eff}$ is the effective action for $\phi$. Now, the partition function is dominated by the external $S_{eff}$ which means we can obtain the leading $\phi$ as 

\begin{equation}
    \begin{split}
        \frac{\delta S_{eff}}{\delta \phi}&=0 \Longrightarrow i\phi=2U \text{Tr} \frac{1}{\partial_\tau+\epsilon_{p}\delta_{ij}-i\phi-\frac{U}{2}-\frac{K_0^2}{\partial_\tau \delta_{ij}-t_{ij}}}
        \label{menai-equation}
    \end{split}
\end{equation}

Now, we introduce the two-point  Green's function for the electrons in the conduction band and the  $p$-electrons 

\begin{equation}
        G_p(\mathbf{k},i\omega_n)= \frac{1}{i\omega_n-\epsilon_p+i\phi(i\omega_n)+\frac{U}{2}-\frac{K_0^2}{i\omega_n-\epsilon_\mathbf{k}}}
        \label{equationSCF=2}
\end{equation}

and

\begin{equation}
    G_c(\mathbf{k},i\omega_n)=\frac{1}{i\omega_n-\epsilon_\mathbf{k}-\frac{K_0^2}{i\omega_n-\epsilon_p+i\phi(i\omega_n)+\frac{U}{2}}}
    \label{GreenF-conduction}
\end{equation}

where $\omega_n=(2n+1)\pi T$ is the fermionic Matsubara frequency, and $\epsilon_\mathbf{k}=-\sum_ie^{-i\mathbf{k\cdot R_{ij}}}t_{ij}$ is the dispersion of free conduction electron. Therefore,  if equation \ref{menai-equation} can be solved self-consistently, one can obtain the Green's function for both $p$-electron and conduction electron.

We then apply the analytical continuation to the self-consistent equations  to obtain the corresponiding self-consistent equation for the retarded Green's function $G_{c,p}^R(\mathbf{k},\omega)$. By solving these equations, we can determine the spectral function averaged over momentum and spin as 

\begin{equation}
    A_{c,p}(\omega)=-\frac{1}{\pi}\int \frac{d^2k}{(2\pi)}\text{Im} \text{Tr} G_{c,p}^R(\mathbf{k},\omega).
    \label{spectadf}
\end{equation}

It is interesting to note that equations \ref{spectadf}  and  \ref{equationSCF=2} can be related to the ones in the dynamical-field theroy (DMFT) for the usual Anderson-Model$^{10}$. It should point out that this theory neglect direct spatial correlation between the $p$-electron, thus only local correlation is included and the self-energy of $p$-electron has not spacial/momentum dependence but this is only the function of time/energy.

\subsection{Non-interacting limit}
When  $U=0$ interaction is turn off, we have a non-interacting system we can obtain the simple two band model 

\begin{equation}
    H_0= -\sum_{\langle i,j\rangle}\sum_{\sigma}t_{ij}c_{\sigma}^\dagger (\mathbf{r}_i) c_{\sigma}(\mathbf{r}_j)+ \epsilon_p\sum_{i=1}^{N}\sum_{\sigma} p_{\sigma}^\dagger (\mathbf{r}_i) p_{\sigma}(\mathbf{r}_i)+\sum_{i=1}^N\sum_{\sigma}
 U_{imp}(\mathbf{r}_i) \left(c_{\sigma}^\dagger (\mathbf{r}_i) p_{\sigma}(\mathbf{r}_i)+ p_{\sigma}^\dagger (\mathbf{r}_i) c_{\sigma}(\mathbf{r}_i)\right)
\end{equation}

Then, the free Green's function are found to be 

\begin{equation}
    G_c(\mathbf{k},i\omega_n)=\frac{1}{i\omega_n-\epsilon_\mathbf{k}-\frac{K_0^2}{i\omega_n-\epsilon_p}}
    \label{GreenF-conduction}
\end{equation}

and 
\begin{equation}
        G_p(\mathbf{k},i\omega_n)= \frac{1}{i\omega_n-\epsilon_p-\frac{K_0^2}{i\omega_n-\epsilon_\mathbf{k}}}
\end{equation}

With these simple equation,  we can calculate the DOS for $p$-electron and conduction bands which have  simple form 

\begin{equation}
    \rho_p(\omega)=-\frac{1}{\pi} \int \frac{d^2k}{(2\pi)^2} \text{Im}G_p(\mathbf{k},\omega)= \frac{K_0^2}{(\omega-\epsilon_p)^2}\int \frac{d^2k}{(2\pi)^2}\delta \left(\omega-\epsilon_\mathbf{k}^c-\frac{K_0^2}{\omega-\epsilon_p} \right)
\end{equation}

and 

\begin{equation}
        \rho_c(\omega)=-\frac{1}{\pi} \int \frac{d^2k}{(2\pi)^2} \text{Im}G_c(\mathbf{k},\omega)=\int \frac{d^2k}{(2\pi)^2}\delta \left(\omega-\epsilon_\mathbf{k}^c-\frac{K_0^2}{\omega-\epsilon_p} \right). 
\end{equation}

Here, we can show that total energy dispersion for a generalized honeycomb lattice for the conduction (valence) band is given by$^{11}$

\begin{equation}
    \epsilon_\mathbf{k}^{c/v}=\sum_ie^{i\mathbf{k\cdot \mathbf{R}_{ij}}}t_{ij}=\pm \sqrt{t_a^2+t_b^2+t_c^2+2t_at_b\cos \left(\frac{\sqrt{3}}{2}ak_x-\frac{1}{2}ak_y  \right)+2t_at_c\cos \left(\frac{\sqrt{3}}{2}ak_x+\frac{1}{2}ak_y  \right)+2t_bt_c\cos(ak_y)}
\end{equation}

which it only considers the nearest-neighbor hopping. Here, $a=3.86$ $\AA$ is the lattice constant for silicene. The above expression can be reduced by  setting  $t_a=t_b=t_c=$  and expanding  the energy dispersion  around the $K$-point, i.e. $\epsilon_\mathbf{k}=\hbar v_F |\mathbf{k}|$. 

\subsection{Weak coupling limit}
Next, if the interaction is weak compered  to non-interacting bands, we can use perturbation theory to extract physical information reliably. For our model, the first-order self-energy is obtained by using the self-consistent Born approximation at zero temperature, which is given by

\begin{equation}
    \text{Re}[i\phi^{(1)}(\omega)]\sim 2 U \pi  \frac{K_0^2}{(\omega-\epsilon_p+\frac{U}{2})^2} N \left(\omega-\frac{K_0^2}{\omega-\epsilon_p+\frac{U}{2}}\right)
\end{equation}

where $N(\omega)\sim \frac{\omega}{\pi(\hbar v_F)^2}$. Therefore, at weak coupling limit, we can write the $p$-electron and conduction Green's functions as 

\begin{equation}
    G_p(\mathbf{k},\omega,U)= \frac{1}{i\omega_n-\tilde{\epsilon_p}-\frac{K_0^2}{i\omega_n-\epsilon_\mathbf{k}}+\frac{2U}{(\hbar v_F)^2} \frac{K_0^2}{(i\omega_n-\tilde{\epsilon_p})^2}\left(i\omega_n-\frac{K_0^2}{i\omega_n-\tilde{\epsilon_p}}\right)}
\end{equation}

and 

\begin{equation}
    G_c(\mathbf{k},i\omega,U)= \frac{1}{i\omega_n-\epsilon_\mathbf{k}-\frac{K_0^2}{i\omega-\tilde{\epsilon_p}+\frac{2U}{(\hbar v_F)^2} \frac{K_0^2}{(i\omega_n-\tilde{\epsilon_p})^2}\left(i\omega_n-\frac{K_0^2}{i\omega_n-\tilde{\epsilon_p}}\right)}}. 
\end{equation}

\begin{figure}[ht]
  \includegraphics[width=17cm]{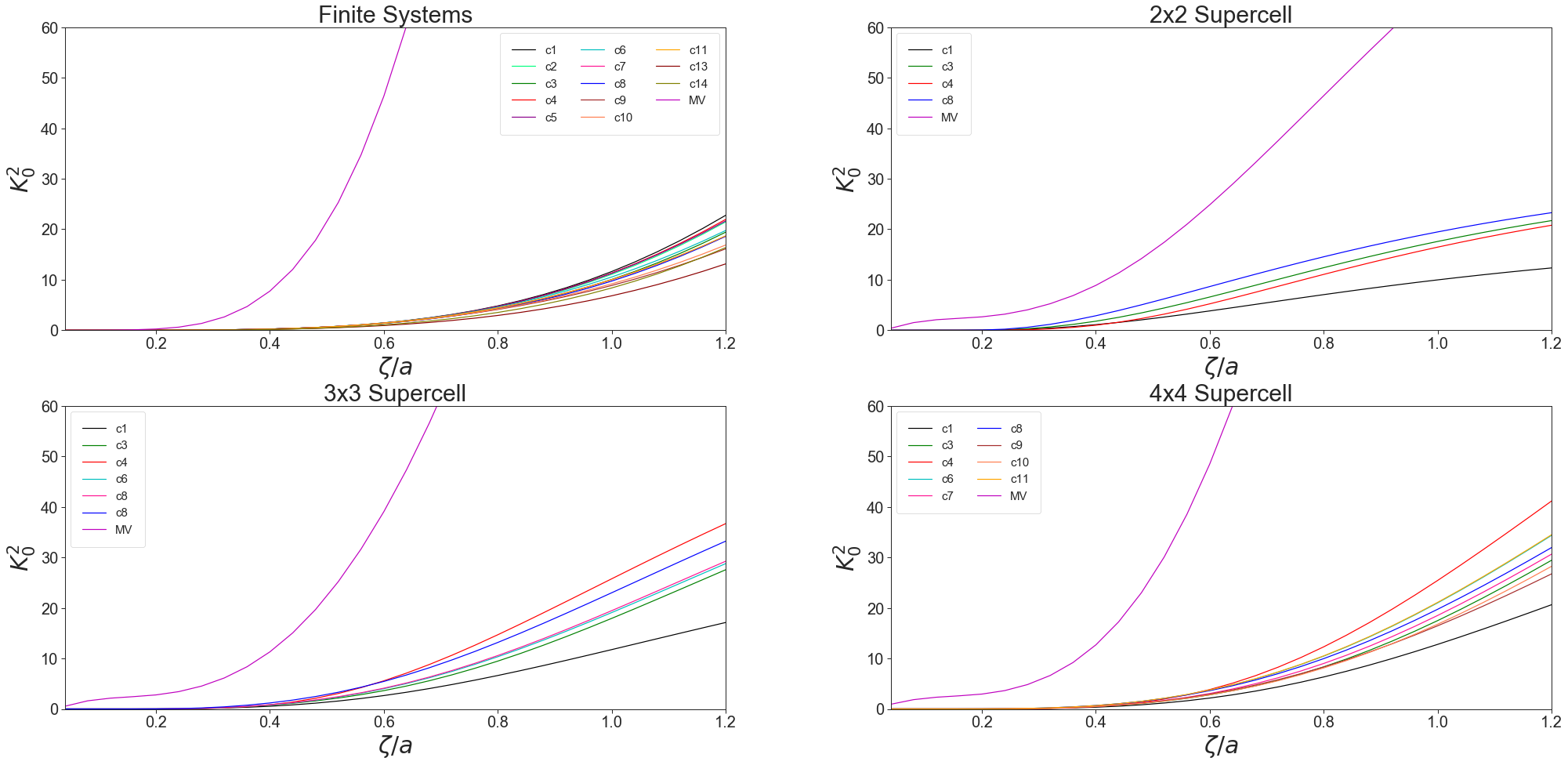}
  \caption{Dimensionnaless correlater for the  C-doped silicene structures.}
  \label{fig:k0_dis}
\end{figure}

where $\tilde{\epsilon_p}=\epsilon_p-\frac{U}{2}$. The respective density of states for the conduction and $p$-electron are given by 

\begin{equation}
    \rho_p (\omega)= -\frac{1}{\pi}\int \frac{d^2k}{(2\pi)^2}\text{Im} G_p (\mathbf{k},\omega,U)
\end{equation}

and 

\begin{equation}
    \rho_c (\omega)= -\frac{1}{\pi}\int \frac{d^2k}{(2\pi)^2}\text{Im} G_c (\mathbf{k},\omega,U)
\end{equation}

Furthermore, $U$ can be positive (repulsive interactions) or negative (attraction interactions) depending of the system of study$^{12}$.

\section{Disorder strength for the C-doped silicene structures}

Figure \ref{fig:k0_dis} shows the dimensionless correlator versus the correlation length for both periodic and finite structures. For the periodic structures, it seems like those structures in which the dopping atoms locate at different sublattices introduce a higher disorder than those located at the same sublattice. The extreme case being the position $S_4$ in which both doping atoms are bonded; however, as the distance for the C atoms increases this trend seems to reverse, as $S_{10} < S_{11}$ in terms of $K_0$. For the finite case, structures that involve a doping site near the center of the cluster tend to have a higher disorder; on the contrary, those structures away from the center decrease the disorder introduced into the layer.

\begin{center}
{\bf References}
\end{center}

\begin{enumerate}
\item Luan, H.-x.;  Zhang, C.-w.;  Li, F.;  Wang, P.-j. Electronic and magnetic properties ofsilicene nanoflakes by first-principles calculations, \textit{Physics Letters A} {\bf 377}, 2792–2795 (2013)
\item Lieb, E. H. Two theorems on the Hubbard model, \textit{Phys. Rev. Lett} {\bf 62}, 1201–1204 (1989)
\item Kresse,  G.;  Hafner,  J.  Ab  initio  molecular-dynamics  simulation  of  the  liquid-metal–amorphous-semiconductor  transition  in  germanium, \textit{Phys.  Rev.  B} {\bf 49},  14251–14269 (1994)
\item Coury, M.; Dudarev, S.; Foulkes, W.; Horsfield, A.; Ma, P.-W.; Spencer, J. Hubbard-like Hamiltonians for interacting electrons in s,  p,  and d orbitals, \textit{Physical  Review  B} {\bf 93}, 075101 (2016)
\item Ulmke, M.;  Janiˇs, V.;  Vollhardt, D. Anderson-Hubbard model in infinite dimensions, \textit{Physical Review B} {\bf 51}, 10411 (1995)
\item Paez. J.; DeLello, K.; Le, D.; Pereira, A. L.; Mucciolo, E. R. Disorder effect on the anisotropic  resistivity  of  phosphorene  determined  by  a  tight-binding  model, \textit{Physical Review B} {\bf 94}, 165419 (2016)
\item Katsnelson, M. Minimal conductivity in bilayer graphene, \textit{The European Physical Journal B-Condensed Matter and Complex Systems} {\bf 52}, 151–153 (2006)
\item Dotsenko, V.Introduction to the replica theory of disordered statistical systems; Cam-bridge University Press, 2005; Vol. 4.
\item Hirsch, J. E. Discrete Hubbard-Stratonovich transformation for fermion lattice models, \textit{Physical Review B} {\bf 28}, 4059 (1983)
\item Byczuk, K.; Hofstetter, W.; Vollhardt, D. Competition between Anderson localizationand  antiferromagnetism  in  correlated  lattice  fermion  systems  with  disorder, \textit{Physical Review letters} {\bf 102}, 146403 (2009)
\item  Hasegawa, Y.; Konno, R.; Nakano, H.; Kohmoto, M. Zero modes of tight-binding elec-trons on the honeycomb lattice, \textit{Physical Review B} {\bf 74}, 033413 (2006)
\item Mancini,  F.;  Marinaro,  M.;  Matsumoto,  H.  Some  properties  of  the  positive-U  andnegative-U hubbard model, \textit{International Journal of Modern Physics B} {\bf 10}, 1717–1734 (1996)
\end{enumerate}

\end{document}